\documentstyle[preprint,aps,amssymb,epsf]{revtex}  
\tightenlines

\begin{document}
\draft                           


\title{Spintronics?}                           

\author{M.I. Dyakonov}  

\address{Laboratoire de Physique Math\'ematique,  Universit\'e 
         Montpellier 2, place E. Bataillon, 34095 Montpellier, France. E-mail: dyakonov@lpm.univ-montp2.fr}


\maketitle


\begin{abstract}

This is a brief review of spin physics in semiconductors, as well as of the historic roots of the recent very active research of spin-related phenomena. The perspectives of "spintronics" are also discussed.

\end{abstract}

---------------------------------------------------------------

\vskip 2cm

{\bf 1. Introduction}\\

 During several last years there has been an explosive development of research in spin physics in semiconductors - research that indeed has yielded a large variety of interesting and spectacular phenomena.  Since this is not sufficient for fund-raising purposes, virtually every article describing this kind of research presents the following justification: this work is important for future quantum computation as well as for the emerging field of spintronics.
  
My view of the perspectives of practical quantum computing has already been published in the preceding Future Trends in Microelectronics collective treatise.\cite{1} As to spintronics, it has been broadly defined as  "...spin-based electronics, where it is not the electron charge but the electron spin that carries information, and this offers opportunities for a new generation of devices combining standard microelectronics with spin-dependent effects..." \cite{2}  This definition seems somewhat too broad, since it includes, for example, magnetic recording and NMR imaging both known  long  before the term "spintronics" was born. What most people apparently mean by spintronics is the fabrication of some useful devices using a) creation of a non-equilibrium spin density in a semiconductor, b) manipulation of the spins by external fields, and c) detection of the resulting spin state.  Narrowing the subject in this way, I will not consider phenomena where the spins are in thermal equilibrium, such as the famous Giant Magnetoresistance Effect, which has already found important applications.
The purpose of this chapter is to briefly review spin physics in semiconductors, as well as the historic roots of recent research.  The real achievements to date and the perspectives for future developments will be discussed.

\vskip 2cm
	
{\bf 2. Spin orientation. Historical background}\\

Maybe the first step towards today's activity was made by Robert Wood in 1923/24 when even the notion of electron spin was not yet introduced.  In a charming paper \cite{3}  Wood and Ellett describe how the initially observed high degree of polarization of mercury vapor fluorescence (resonantly excited by polarized light) was found to diminish significantly in later experiments.  "It was then observed that the apparatus was oriented in a different direction from that which obtained in earlier work, and on turning the table on which everything was mounted through ninety degrees, bringing the observation direction East and West, we at once obtained a much higher value of the polarization." In this way Wood and Ellet discovered what we now know as the Hanle effect, \cite{4} i.e. depolarization of luminescence by transverse magnetic field (the Earth's field in their case). It was Hanle who carried out detailed studies of this phenomenon and provided the physical interpretation.

	The subject did not receive much attention until 1949 when Brossel and Kastler\cite{5} initiated profound studies of optical pumping in atoms, which were conducted by Kastler and his school in Paris in the 50ies and 60ies. (See Kastler's Nobel Prize award lecture.\cite{6}) The basic physical ideas and the experimental technique of today's "spintronic" research originate from these seminal papers: creation of a non-equilibrium distribution of atomic angular moments by optical excitation, manipulating this distribution by applying $dc$ or $ac$ fields, and detecting the result by studying the luminescence polarization.  The relaxation times for the decay of atomic angular moments can be quite long, especially when hyperfine splitting due to the nuclear spin is involved. 

	A number of important applications have emerged from these studies, such as gyroscopes and hypersensitive magnetometers, but in my opinion, the  knowledge obtained is even more valuable.  The detailed understanding of various atomic processes and of many aspects of interaction between light and matter was pertinent to the future developments, e.g. for laser physics.

	The first experiment on optical spin orientation of electrons in a semiconductor (Si) was done by Georges Lampel\cite{7} in 1968, as a direct application of the ideas of optical pumping in atomic physics.  The important difference is that now these are the free conduction band electrons that get spin-polarized, rather then electrons bound in an atom.  This pioneering work was followed by extensive experimental and theoretical studies mostly performed by small research groups at Ecole Polyt\'echnique in Paris and at Ioffe Institute in St Petersburg (Leningrad) in the 70ies and early 80ies . At the time this research met with almost total indifference by the rest of the physics community.   This lack of interest can be partly explained by the fact that no obvious applications were in sight, and the word "spintronics"" was not yet invented.

\vskip 2cm
	
{\bf 2.  Basic spin interactions }\\
 
This section serves to enumerate the possible types of spin interactions that can be encountered in a semiconductor.

{\it 3.1. Magnetic interaction.} This is the direct dipole-dipole interaction between the magnetic moments of a pair of electrons.  For two electrons located at neighboring sites in a crystal lattice this gives energy on the order of 1K.  This interaction is normally too weak to be of any importance in semiconductors.\\

{\it 3.2.  Spin-orbit interaction.}  If an observer moves with a velocity ${\bf v}$ in an external electric field ${\bf E}$, he will see a magnetic field ${\bf B} = ({\bf v}/c)\times {\bf E}$, where $c$ is the velocity of light  This magnetic field acts on the electron magnetic moment. This is the physical origin of the spin-orbit interaction, the role of which strongly increases for heavy atoms (large $Z$). Due to spin-orbit interaction, any electric field, whether static or alternating, acts on the spin of a moving electron.
   
	Spin-orbit interaction is key to our discussion as it enables optical spin orientation and detection (the electrical field of the light wave does not interact directly with the electron spin). It is (in most cases) responsible for spin relaxation. And finally, it makes the transport and spin phenomena inter-dependent.\\

 {\it 3.3. Exchange interaction.} It is, in fact, the result of the electrostatic Coulomb interaction between electrons, which becomes spin-dependent because of the requirement of the Pauli exclusion principle. The exchange interaction is responsible for ferromagnetism and for the phenomenon of spin injection in a semiconductor through a ferromagnetic contact.  Because of low free electron density in a semiconductor, compared to a metal, it is normally not of major importance, except magnetic semiconductors (like CdMnTe) and the semiconductor-ferromagnet interface.\\ 

{\it 3.4. Hyperfine interaction with nuclear spins.}  This is the magnetic interaction between the electron and nuclear spins, which may be quite important if the lattice nuclei in a semiconductor have none-zero spin (like in GaAs). If the nuclei get polarized, this interaction is equivalent to the existence of effective nuclear magnetic field acting on electron spins. The effective field of 100\% polarized nuclei in GaAs would be several Tesla!  Experimentally, nuclear polarization of several percent is easily achieved.

\vskip 2cm

{\bf 4.  Brief review of spin physics in semiconductors}\\

The basic ideas concerning spin phenomena in semiconductors were developed both theoretically and experimentally 30 years ago.  Some of these ideas have been rediscovered only recently. A review of non-equilibrium spin physics in (bulk) semiconductors can be found in Ref. \cite{8}, as well as in other chapters of the Optical Orientation book.\\ 

{\it 4.1. Optical spin orientation and detection.}  To date, the most efficient way of creating non-equilibrium spin orientation in a semiconductor is provided by interband absorption of circularly polarized light.  Photons of right or left polarized light have a projection of the angular momentum on the direction of their propagation equal to $+1$ or $-1$, respectively.  When a circularly polarized photon is absorbed, this angular momentum is distributed between the photoexcited electron and hole according to the selection rules determined by the band structure of the semiconductor.  Because of the complex nature of the valence band, this distribution depends on the value of the momentum of the created electron-hole pair ($+{\bf p}$ and $-{\bf p}$).  However if we take the average over the directions of ${\bf p}$, the result is the same as in optical transitions between atomic states with $j=3/2$ (corresponding to bands of light and heavy holes) and $j=1/2$ (corresponding to the conduction band).  The photocreated electrons get spin-polarized because of the strong spin-orbit interaction in the valence band.

In an optical spin orientation experiment a semiconductor (usually {\it p}-type) is excited by circularly polarized light with $\hbar \omega>E_g$.  The circular polarization of the luminescence is analyzed, which gives a direct measure of the electron spin polarization (actually, the degree of circular polarization is simply equal to the average electron spin).  Thus various spin interactions can be studied by simple experimental means.  The electron polarization will be measured provided that the spin relaxation time $\tau_s$ is not very short compared to the recombination time $\tau$, a condition, which often can be achieved even at room temperature. \\

{\it 4.2. Spin relaxation.} Spin relaxation, i.e. disappearance of initial non-equilibrium spin polarization, is the central issue for all spin phenomena.  (In recent years it has become fashionable to use the term "spin decoherence" for this process). Spin relaxation can be generally understood as a result of the action of fluctuating in time magnetic fields. In most cases, these are not "real" magnetic fields, but rather "effective" magnetic fields originating from the spin-orbit, or, sometimes, exchange interactions (see Section 1). A randomly fluctuating magnetic field is characterized by two important parameters: its amplitude (or, more precisely, its rms value), and its correlation time $\tau_{c}$, i.e. the time during which the field may be roughly considered as constant.  Instead of the amplitude, it is more convenient to use the average spin precession frequency in this random field, $\omega$.
  
Thus we have the following physical picture of spin relaxation:  the spin makes a precession around the (random) direction of the effective magnetic field with a typical frequency $\omega$ and during a typical time $\tau_{c}$.  After a time $\tau_{c}$ the direction and the absolute value of the field change randomly, and the spin starts its precession around the new direction of the field.  After a certain number of such steps the initial spin direction will be completely forgotten. 
 
What happens depends on the value of the dimensionless parameter $\omega \tau_{c}$, which is the typical angle of spin precession during the correlation time.  There are two limiting cases  to be considered. If $\omega \tau_{c}<<1$, which is usualy the case,  the precession angle is small, so that the spin vector experiences a slow angular diffusion.  During a time interval $t$, the number of random steps is $t/\tau_{c}$, and for each step the squared precession angle is $(\omega \tau_{c})^2$. These steps are not correlated, so that the total squared angle after time $t$ is $(\omega \tau_c)^{2}t/\tau_c$.  The spin relaxation time $\tau_s$ may be defined as the time at which this angle becomes of the order of unity:  $1/\tau_s\sim \omega^{2}\tau_c$.

This is essentially a {\it classical} result (the Planck constant does not enter), although certainly it can be also derived quantum-mechanically.  Note, that for this case  $\tau_s>>\tau_c$.

	In the opposite limit, $\omega \tau_c>>1$, a spin will make many rotations around the direction of the magnetic field during the correlation time.  During the time on the order of $1/\omega$ the spin projection transverse to the random magnetic field is (on average) completely destroyed, while its projection along the direction of the field is conserved.   After time $\tau_c$ the magnetic field changes its direction, and the initial spin polarization will disappear.  Thus for this case $\tau_s\sim \tau_p$, i.e. the spin relaxation time is on the order of the correlation time. 

	There are many possible mechanisms providing the fluctuating magnetic fields responsible for spin relaxation.  For mobile electrons in bulk III-V and II-VI semiconductors, as well as for 2D electrons in quantum wells, the dominant process is usually the Dyakonov-Perel mechanism \cite{9},\cite{10},\cite{11} in which the effective magnetic field is related to the momentum dependent spin splitting of the conduction band.  For electrons localized at donor sites or in quantum dots, this and other mechanisms related to the electron motion do not work, and the spin relaxation times may become quite long. \\

{\it 4.3. Hanle effect.}  Depolarization of luminescence by a transverse magnetic field (first discovered by Wood and Ellett, as described above) is effectively employed in experiments on spin orientation in semiconductors.  The reason for this effect is the precession of electron spins around the direction of the magnetic field. Under continuous illumination, this precession leads to the decrease of the average projection of the electron spin on the direction of observation, which defines the degree of circular polarization of the luminescence.  Thus the degree of polarization decreases as a function of the transverse magnetic field.  Measuring this dependence under steady state conditions makes it possible to determine both the spin relaxation time and the recombination time.\\

{\it 4.4.  Spin-related currents.}  There are several, generally weak effects based on the interconnection between spin and charge transport.  One of them is related to spin-orbit interaction in scattering of electrons, which leads to the so-called skew scattering, an asymmetry of scattering with respect to the plane containing the directions of the initial spin and initial momentum - also known as the Mott effect.  This effect leads to the appearance of an additional electron current, ${\bf j}$, due to the existence of spin orientation and its gradient:$$ $$

\hskip 150pt ${\bf j} = - \beta {\bf E\times S} - \delta $curl $  {\bf S}$,
$$ $$
where  ${\bf S}$ is the spin density vector, ${\bf E}$ is the electric field, and the coefficients $\beta$ and $\delta$ are proportional to the spin-orbit interaction. The first term is responsible for the so-called anomalous Hall effect. The second one describes an additional current component appearing due to an inhomogeneous spin density,\cite{12,13} an effect  that was first reported in Ref. \cite{14}. An inverse effect is also possible: a current flow may create spin orientation near the sample surface.\cite{15}
  
In gyrotropic crystals a current can be induced by a homogeneous non-equilibrium spin density, as it was shown theoretically by Ivchenko and Pikus \cite{16} and by Belinicher.\cite{17}  This first experimental demonstration of this effect was reported in Ref. \cite{18}.\\ 

{\it 4.5.  Interaction between the electron and nuclear spin systems.} The non-equilibrium spin-oriented electrons can easily transmit their polarization to the lattice nuclei, thus creating an effective magnetic field (quite large, up to a fraction of Tesla). This field will, in turn,  influence the spin of electrons (but not on their orbital motion). For example, it can strongly influence the electron polarization via the Hanle effect.  Thus the spin-oriented electrons and the polarized lattice nuclei form a strongly coupled system, in which spectacular non-linear phenomena, like self-sustained slow oscillations and hysteresis are observed by simply looking at the circular polarization of the luminescence. \cite{19,20} Optical detection of the nuclear magnetic resonance in a semiconductor was demonstrated for the first time by Ekimov and Safarov. \cite{21}\\
 
{\it 4.6.  Spin injection.} The idea of creating non-equilibrium electron spins in a semiconductor by passing a current through a ferromagnetic contact was proposed by Aronov and Pikus.\cite{22} The spin orientation on the semiconductor side should persist on the spin diffusion length $(D\tau_s)^{1/2}$, where $D$ is the diffusion coefficient and  $\tau_s$ is the spin relaxation time. Such experiments were performed only recently (see below).
  
\vskip 2cm

{\bf 5. Recent research}\\

Within the scope of this paper it is only possible to briefly outline the main directions of the current research presented in hundreds of publications. Most of this work is focused on spin-related optical and transport properties of two-dimensional semiconductor structures.\\

{\it 5.1. Time-resolved optical experiments.} The time resolved optical measurements technique, developed by Awschalom' and co-workers \cite{23} has allowed to visualize very clearly and impressively the spin precession and the spin relaxation of electrons and holes on the picosecond time scale.  This gives a unique possibility to study the intimate details of various spin processes in a semiconductor.\\ 

{\it 5.2. Spin dynamics in magnetic semiconductors.} Mn doped III-V and II-VI systems, both bulk and two-dimensional, have attracted intense interest.\cite{24} The giant Zeeman splitting due to exchange interaction with Mn, combination of ferromagnetic and semiconductor properties, and the possibility of making a junction between a ferromagnetic and a normal semiconductor have been the focus of numerous studies.\\

{\it 5.3.  Spin dynamics in quantum dots.}  The interest for spin effects in quantum dots is related to the expected very low spin relaxation rates (which are not observed so far, possibly because of non-homogeneous broadening). The only relaxation mechanisms that should remain effective are either  phonon-related or due to the hyperfine interaction with nuclear spins. A number of studies addressed the spin dependence of transport through a quantum dot due to the Pauli eclusion principle.\\ 

{\it 5.4. Spin injection.} Potential barriers at a ferromagnet-semiconductor interface are different for electrons with spin parallel or antiparallel to the polarization in the ferromagnet because of exchange interaction. For this reason, any departure from equilibrium (e.g. due  to electric current through the interface or to illumination) will generally lead to spin injection in the semiconductor.  The non-equilibrium spin orientation will exist on the scale of the spin diffusion length $(D\tau_s)^{1/2}$. Experimentally, spin injection from a ferromagnet to a normal metal, originally proposed by Aronov,\cite{25} was first observed by Johnson and Silsbee.\cite{26} Injection through a ferromagnet/semiconductor junction (with an efficiency of several percent) was investigated in many recent works.\\

{\it 5.5. The Datta-Das spin transistor.} Virtually every paper in the field cites as an example of a future spintronic device the so-called Datta-Das transistor \cite{27}, which is a ballistic field effect transistor with ferromagnetic source and drain contacts.  The source contact injects electrons with a fixed direction of spin, which can be rotated during the time of flight towards the drain by applying a gate voltage. This spin rotation arises due to spin-orbit interaction, with the gate-induced electric field transforming into an effective magnetic field (the Bychkov-Rashba effect \cite{28}). Since the magnetic drain contact preferentially accepts a certain spin direction, the current will be an oscillating function of the gate voltage.

   Datta and Das modestly labeled this proposed device as an "electronic analog of the electro-optic modulator" which  is quite correct.  The main difference  lies in the fact that an electro-optic modulator produces polarization-modulated light which can be used elsewhere, while the electronic analog just measures the modulated spin polarization within a submicron distance from the source. Such a device is yet to be fabricated and faces a number of obstacles, including the existence  of stray magnetic fields from the contacts and the generally low efficiency of spin injection.  However, even if it were fabricated one day, the advantages of such a spin transistor are far from obvious.  After all, it will basically control the drain current by gate voltage - a task that is routinely performed without involving any spin physics.\\   

{\it 5.6. Spin-dependent tunneling.}  If a tunneling contact is ferromagnetic, the potential barriers for spin up and spin down electrons are different. So are the transmission probabilities, and thus electrons with a certain spin direction will be preferentially transmitted.  Injection of spin in a GaAs by electron tunneling from a ferromagnetic tip was demonstrated for the first time by Alvarado and Renaud.\cite{29} Similar phenomena was studied in many recent works. \\

{\it 5.7. Spin-related currents.} Numerous recent results on spin photocurrents (also known as the circular photo-galvanic effect) in two-dimensional structures can be found in the comprehensive review by Ganichev and Prettl. \cite{30} These studies are very interesting and involve subtle physics, which is now quite well understood.\\ 

{\it 5.8. Nuclear spin effects in optics and electron transport.}  As far as optical dynamic nuclear polarization in semiconductors and its optical detection are concerned, no major advances were made recently compared to the early results presented in Refs. \cite{8,20,31}. There are however some spectacular results on nuclear spin effects in low temperature magneto-transport.\cite{32} Enormous changes of the magnetoresistance in the Fractional Quantum Hall Effect regime are observed and shown to be related to the lattice nuclear spin polarization.  This interesting effect is still waiting for a theoretical interpretation. \\

\vskip 2cm
{\bf 6. Conclusions}\\

The field of spin physics in semiconductors is extremely rich and interesting. There are plenty of spin-related effects in optics and transport. While most of these effects are known for some time and the basics of spin physics are well established, the current very active research in spin dynamics in semiconductor and semi/ferro nanostructures is yielding new experimental techniques, as well as a better understanding of material properties, and revealing new possibilities of creation and detection of non-equilibrium spins. 

This research, which is certainly interesting and exciting by itself, is being "sold" under the banner of  "spintronics", a future powerful branch of electronics where the electron spin will be used for more efficient information processing {\it etc}.  For this end, a number of fairly vague ideas concerning spin-based devices have been advanced. 
 Are there any real grounds for such a vision?  Certainly, the key aspect of the future is that it is unpredictable.  One may recall many examples of high expectations (such as "all-optical computer" or "Josefson junction logic"), which were later abandoned. Also worth noting is that several decades of intense research of NMR were required before really important practical applications appeared.
  
My opinion is that so far there are no indications of a coming spintronic revolution in electronics, nor any clear enough ideas of how it could happen, although it is quite probable that some applications will emerge. The electron spin is analogous to light polarization: while some applications of polarization effects in optics do exist, they clearly are not of primary practical importance. The difference between charge and spin is that once a voltage is applied, charges are likely to flow, leading to a usefull current.  Non-equilibrium spin orientation, on the other hand, is even more fragile than light polarization, and the requirement of long spin relaxation times generally contradicts the possibility of easy spin manipulation.

It is amusing to read an increasing number of theoretical articles, which claim some weak and exotic spin effect to be of great importance for future spintronics and/or quantum computing. Clearly, the existence of some spin-dependent property does not necessarily mean that it can or will be used in practice. 

In summary, while high-quality research in spin physics, like in any other domain, is surely beneficial, it still remains to be seen whether it will result in any kind of breakthrough in electronics.


\end{document}